\documentclass[aps,prc,onecolumn,floatfix,superscriptaddress]{revtex4-2}
\usepackage{blindtext}
\usepackage{float}
\usepackage{amsmath}
\usepackage{amssymb}
\usepackage{graphicx}
\usepackage [english]{babel}
\usepackage{todonotes}
\usepackage{braket}

\begin{document}

\def\nuc#1#2{\relax\ifmmode{}^{#1}{\protect\text{#2}}\else${}^{#1}$#2\fi}
\title{A formalism to assess the accuracy of nuclear-structure weak interaction effects in precision $\beta$-decay studies}

\author{Ayala Glick-Magid}
\affiliation{The Racah Institute of Physics, The Hebrew University of Jerusalem, Givat Ram, Jerusalem, 9190401}
\author{Doron Gazit}
\email[E-mail:~]{doron.gazit@mail.huji.ac.il}
\affiliation{The Racah Institute of Physics, The Hebrew University of Jerusalem, Givat Ram, Jerusalem, 9190401}

\begin{abstract}
Multiple high precision $\beta$-decay measurements are being carried out these days on various nuclei, in search of beyond the Standard Model signatures.
These measurements necessitate  accurate standard model theoretical predictions to be compared with.
Motivated by the experimental surge, we present a formalism for such a calculation of $\beta$-decay observables, with controlled accuracy, based on a perturbative analysis of the theoretical observables related to the phenomena, including high order nuclear recoil and shape corrections. The accuracy of the corrections is analyzed by identifying a hierarchy of small parameters, related to the low momentum transfer characterizing $\beta$-decays. 
Furthermore, we show that the sub-percent uncertainties, targeted by on-going and planned experiments, entail an accuracy of the order of 10\% for the solution of the nuclear many body problem, which is well within the reach of modern nuclear theory for light to medium mass nuclei.
\end{abstract}
\maketitle

\section{Introduction}

According to the Standard Model (SM), the weak interaction of quarks and weak probes has a $V-A$ structure \cite{1742-6596-196-1-012002,RevModPhys.78.991}, i.e., a polar-vector minus an axial-vector current, with the same magnitude but opposite signs. This has been experimentally tested in many ways, including by studying correlations in the directions of the outgoing leptons in nuclear $\beta$-decays \cite{PhysRev.105.1413,PhysRev.106.517}. However, there are pieces of evidence showing that the SM is incomplete, even much below the Planck scale, e.g., the existence of the neutrino mass. Since the neutrino, as a low mass neutral lepton, interacts almost exclusively via the weak force, this might hint that the way we understand the weak interaction is incomplete.

These days, angular correlation and spectrum measurements in nuclear
$\beta$-decays are carefully designed to probe for signatures of deviations
of the weak interaction from the $V-A$ symmetry dictated by the SM, originating in interactions with scalar, pseudo-scalar,
or tensor symmetries, which are a-priori theoretically allowed in
Lorentz invariant theories \cite{PhysRev.82.531}. Such signature
would be an indication for Beyond the Standard Model (BSM) physics. Dimensional
analysis shows that the order of magnitude of such deviation is inversely
proportional to the square of the new physics energy scale, e.g.,
a 1\% deviation from the SM values of those experimental
observables suggests new physics at the TeV-scale. Thus, such measurements
can be viewed as low-energy counterparts for LHC measurements.
New $\beta$-decay experiments are designed to have a per-mill level accuracy~\cite{cirgiliano2019precision}. 
Such new efforts include, for example, studying the decays of $^{6}\text{He}$,
$^{16}\text{N}$, and $^{17-23}\mbox{Ne}$ isotopes at SARAF accelerator (Israel) \cite{refId0,Ohayon2018}, $^{6}\text{He}$ decay studies at the National
Superconducting Cyclotron Laboratory (USA)
\cite{Huyan2016,HUYAN2018134,doi:10.1063/1.4955362},
Laboratoire Physique Corpusculaire de CAEN (France)~\cite{cirgiliano2019precision},
 the University of Washington CENPA (USA) by the He6-CRES Collaboration~\cite{PhysRevLett.114.162501},
and more (for an overview of on-going and planned experiments see \cite{GONZALEZALONSO2019165,cirgiliano2019precision}).
However, deviations from the simplistic textbook formulae are also
due to finite momentum transfer, radiative corrections, and nuclear
structure effects (see e.g., \cite{RevModPhys.90.015008,HAYEN2019152}
and references therein). Pin-pointing these effects demands a detailed
calculation of the nuclear dynamics of the weak decay.
In this paper we present a general formalism, to calculate
the nuclear matrix element of these observables, within the Standard
Model, including e.g., recoil and (spectrum-)shape.
The main advantage of the presented formalism is the possibility to assess a controlled theoretical accuracy, based on an identification
of the parameters governing this accuracy.

This paper is structured as follows: 
We begin by identifying small parameters involved in the $\beta$-decay formalism (sec. \ref{Sec:small_parameters}). 
Next, we introduce the $\beta$-decay multipole expansion, analyze its properties (sec. \ref{sec: formalism}), and employ it to formulate a general perturbative expansion of the observables, including high order corrections, for any $\beta$-decay transition (sec. \ref{sec: general}). 
Following that, we demonstrate the corrections for allowed $\beta$-decays (sec. \ref{sec: allowed}), and for first forbidden decays (sec. \ref{sec: 1fobidden}).
Finally, we discuss the expected order of magnitude of BSM signatures, then consider the accuracy required for SM calculations to detect BSM deviations in nuclear $\beta$-decay measurements (sec. \ref{sec: BSM Signatures}), and summarize (sec. \ref{sec: summary}).

\section{Small parameters governing nuclear structure effects of $\beta$-decays of finite nuclei} \label{Sec:small_parameters}
The differential distribution of a $\beta$-electron of energy $\epsilon$,
momentum $\vec{k}$ and direction $\vec{\beta}=\frac{\vec{k}}{\epsilon}$,
and a neutrino $\nu$ of momentum $\vec{\nu}$, in a $\beta^{\mp}$-decay
process, can be written as \cite{PhysRev.104.254}:
\begin{align}
\label{eq:general-decay-rate}
\frac{d^{5}\omega}{d\epsilon\frac{d\Omega_{k}}{4\pi}\frac{d\Omega_{\nu}}{4\pi}} & =\frac{4}{\pi^{2}}\left(\epsilon_0-\epsilon\right)^{2}k\epsilon F^{\mp}\left(Z_{f},\epsilon\right)C_{\text{corr.}}\frac{1}{2J_{i}+1}\Theta\left(q,\vec{\beta}\cdot\hat{\nu}\right)\text{.}
\end{align}
Here $\left(\epsilon_0,\vec{q}\right)=\left(\epsilon,\vec{k}\right)+\left(\nu,\vec{\nu}\right)$ is the momentum
transfer in the process, i.e., the difference between the initial
momentum and final momentum of the nuclear
states. $J_{i}$ is the total angular momentum of the decaying nucleus, and
$Z_{f}$ is the charge of the nucleus after the decay. The deformation of the beta particle wave function, due to
the long-range electromagnetic interaction with the nucleus, is taken
into account at the leading order by the Fermi function $F^{\pm}\left(Z_{f},\epsilon\right)$
for a $\beta^{\pm}$-decay.

Corrections which do not originate purely in the weak matrix element,
such as radiative corrections, finite mass and electrostatic finite
size effects, and atomic and chemical effects, are represented in
eq. \eqref{eq:general-decay-rate} by $C_{\text{corr.}}$.
As these corrections are assumed known in the literature (see, e.g., \cite{RevModPhys.90.015008}), we will
not focus on them, but on those effects originating from the fact that the weak interaction of a probe with a finite nucleus is affected
by the complex nuclear structure of the decaying and final nuclei.
These are absorbed inside $\Theta\left(q,\vec{\beta}\cdot\hat{\nu}\right)$.
Current experimental efforts are divided into spectrum (shape) measurements or angular correlation measurements \cite{cirgiliano2019precision},
both affected by the nuclear-structure.
Our discussion will distinguish between these corrections.

In order to compare the experimental results to the theory, there
is a need for a good understanding of the theoretical framework. As
the momentum transfer $q$ is limited in $\beta$-decays, usually,
up to few tens MeV/c, typically up to 10 MeV/c, and small compared
to other energy or momentum scales in the nuclear problem, this framework
is simplified by an expansion in a set of small parameters. For example,
$\epsilon_{qr}\equiv qR$ (for an endpoint of $\approx2\text{MeV}$,
$\epsilon_{qr}\approx0.01A^{\frac{1}{3}}$, where $A$ is the mass
number of the nucleus), with $R$ the nucleus's nuclear radius.
In addition, the non-relativistic character of the nuclear problem
introduces,
\begin{align}
\epsilon_{\text{NR}} & \equiv\frac{P_{\text{Fermi}}}{m_{N}}\approx 0.2,\nonumber \\
\epsilon_{\text{recoil}} & \equiv\frac{q}{m_{N}},
\end{align}
where $P_{\text{Fermi}}$ is the Fermi momentum, and $m_{N}$ is the
nucleon mass. These are the non-relativistic expansion small parameter
$\epsilon_{\text{NR}}$, and the recoil correction $\epsilon_{\text{recoil}}$
(e.g., for an endpoint of $\approx 2$ MeV, $\epsilon_{\rm recoil}\approx 0.002$),
both related to the low energy assumption taken in the calculation of the currents.

A different source of deviation from the simplistic $\beta$-decay formulas arises from the Coulomb force between the $\beta$-particle and the nucleus, which deforms the wave function of the $\beta$-particle from a plane-wave~\cite{armstrong1972coulomb}. This results in corrections dominated by $\epsilon_c\equiv\alpha Z_{f}$ \cite{wilkinson1970evaluation}, where $\alpha \approx \frac{1}{137}$ is the fine structure constant.
Those can be divided, as described in Ref.~\cite{holstein1979electromagnetic}, into two:
corrections of the spectrum, reflected in the Fermi function and in additional spectrum corrections whose precise form depends upon the specific $\beta$-transition, and corrections to the recoil form factors.
The latter can also be separated into two: corrections reminiscent of the spectrum corrections, and a correction of the electron energy, following the presence of the Coulomb potential.

The Fermi function is considered known and we will not discuss it here. The non-energy corrections to the form factors, are of $\epsilon_c \epsilon_{qr} \epsilon_{\rm recoil}$ order of magnitude~\cite{armstrong1972coulomb}, and therefore negligible at this stage. The remaining corrections are those that depend on the transition, and the correction of the electron energy.

The correction of the electron energy adds the Coulomb displacement energy to the maximal electron energy \cite{behrens1982electron} appearing in the multipole operators (see appendix). This Coulomb displacement energy is usually taken by its approximation for a homogeneously charged sphere $\Delta E_c \approx \frac{6}{5}\frac{\alpha Z_f}{R}$ \cite{behrens1982electron}, but this is a very crude approximation, and for comparison with experimental measurements we suggest using the experimental value, that can be found in the literature for most nuclei (see, e.g., \cite{antony1997coulomb}).

The last type of Coulomb corrections is the spectrum corrections which are unique to each $\beta$-transition. Those are of the order of $\epsilon_c ^2$ or $\epsilon_c \epsilon_{qr}$~\cite{holstein1974electromagnetic}. While for particularly light nuclei these are negligible, for light-medium nuclei this is already a substantial contribution and there is a need for neat work summarizing them for the different transitions in a form useful for {\it Ab~initio} calculations.
As current experimental efforts focus on light nuclei, we postpone a more detailed discussion in these corrections to future work (see Refs. \cite{behrens1969numerical, behrens1982electron}, and Refs. \cite{RevModPhys.90.015008, hayen2020consistent} for allowed transitions).

The nuclear model is a source of uncertainty, as it is a low energy effective model of QCD. It is generally solved in a few steps. First, a model is written to describe the Hamiltonian and currents representing the quantum process in the nuclear regime. There are several approaches to cope with this problem. A modern one is based on the effective field theory (EFT) approach. EFT expansion relies on scale separation in the low energy nuclear momentum-energy. As $\beta$-decays and nuclei are characterized by low-energies compared to QCD scales, they can be described effectively by an EFT which includes nucleons and pions, an EFT named chiral EFT. EFT is a systematic expansion of the Hamiltonian of the fundamental theory, order-by-order, in a small parameter $\epsilon_{\text{EFT}}=Q/\Lambda_b$ \cite{cirgiliano2019precision}, where $\Lambda_b$ is the breakdown scale of the EFT. In chiral EFT, there is more than just the nucleons and pions as effective degrees of freedom, it is combined with a specific approach to non-relativistically expand the nuclear theory, a specific cutoff regularization and more. We note that this procedure includes a systematic uncertainty, $\epsilon_{\text{EFT}}^n$, where $n$ is the order of the EFT expansion. However, there is an additional systematic error in doing so, $\epsilon_{\text{model}}$. Finally, the resulting equations are solved numerically, introducing a third source of uncertainty, the convergence error $\epsilon_{\text{conv}}$. The combination of all these is denoted by $\epsilon_{\text{NM}}$ (NM is nuclear model). In recent years it has been shown that this procedure reaches $\epsilon_{\text{NM}}$ of a few percents \cite{Gysbers2019}.

In the following, we trace the scaling of different nuclear structure related corrections to the aforementioned dimensionless small parameters.

\section{$\beta$-decay multipole expansion formalism}\label{sec: formalism}

Assuming the Standard Model $V-A$ coupling, the function $\Theta\left(q,\vec{\beta}\cdot\hat{\nu}\right)$ depends on the nuclear wave functions, and is conveniently written using a multipole expansion \cite{Walecka:819209},
\begin{multline}
\Theta\left(q,\vec{\beta}\cdot\hat{\nu}\right)=\sum_{J=1}^{\infty}\left[\left(1-\left(\hat{\nu}\cdot\hat{q}\right)\left(\vec{\beta}\cdot\hat{q}\right)\right)\left(\left|\left\langle \left\Vert \hat{E}_{J}\right\Vert \right\rangle \right|^{2}+\left|\left\langle \left\Vert \hat{M}_{J}\right\Vert \right\rangle \right|^{2}\right)\pm\hat{q}\cdot\left(\hat{\nu}-\vec{\beta}\right)2\mathfrak{Re}\left(\left\langle \left\Vert \hat{E}_{J}\right\Vert \right\rangle \left\langle \left\Vert \hat{M}_{J}\right\Vert \right\rangle ^{*}\right)\right]+\\
+\sum_{J=0}^{\infty}\left[\left(1-\vec{\beta}\cdot\hat{\nu}+2\left(\hat{\nu}\cdot\hat{q}\right)\left(\vec{\beta}\cdot\hat{q}\right)\right)\left|\left\langle \left\Vert \hat{L}_{J}\right\Vert \right\rangle \right|^{2}+\left(1+\vec{\beta}\cdot\hat{\nu}\right)\left|\left\langle \left\Vert \hat{C}_{J}\right\Vert \right\rangle \right|^{2}-\hat{q}\cdot\left(\hat{\nu}+\vec{\beta}\right)2\mathfrak{Re}\left(\left\langle \left\Vert \hat{L}_{J}\right\Vert \right\rangle \left\langle \left\Vert \hat{C}_{J}\right\Vert \right\rangle ^{*}\right)\right]\text{,}\\
\label{eq: full leptonic multipole decomposition-2}
              \end{multline}
where $\left\langle \left\Vert \hat{O}_{J}\right\Vert \right\rangle $ is the reduced matrix element of a rank $J$ spherical tensor operator $\hat{O}_{J}$, between the initial- and final-nuclear state wave functions. We use a multipole operator decomposition of the nuclear current, viz. the Coulomb, longitudinal, electric, and magnetic operators, defined as:
\begin{eqnarray}
\label{eq: multipole operators}
\hat{C}_{J}\left(q\right) & \equiv & \int d^{3}rj_{J}\left(qr\right)Y_{J}\left(\hat{r}\right)J_{0}\left(\vec{r}\right),\nonumber \\
\hat{L}_{J}\left(q\right) & \equiv & \frac{i}{q}\int d^{3}r\left[\vec{\nabla}\left(j_{J}\left(qr\right)Y_{J}\left(\hat{r}\right)\right)\right]\cdot\vec{J}\left(\vec{r}\right),\nonumber \\
\hat{E}_{J}\left(q\right) & \equiv & \frac{1}{q}\int d^{3}r\left[\vec{\nabla}\times\left(j_{J}\left(qr\right)\vec{Y}_{JJ1}\left(\hat{r}\right)\right)\right]\cdot\vec{J}\left(\vec{r}\right),\nonumber \\
\hat{M}_{J}\left(q\right) & \equiv & \int d^{3}rj_{J}\left(qr\right)\vec{Y}_{JJ1}\left(\hat{r}\right)\cdot\vec{J}\left(\vec{r}\right)\mbox{,}
\end{eqnarray}
where $\vec{J}\left(\vec{r}\right)$ and $J_0\left(\vec{r}\right)$ are the nuclear, current and charge, coupling to the probe. $j_{J}$ are the spherical Bessel functions, and $Y_{J}$ and $\vec{Y}_{Jl1}^{M}$ are the spherical harmonics and the vector spherical harmonics, respectively. The nuclear wave functions are commonly calculated numerically using an expansion in a complete basis. In some cases, analytical expressions can be derived, for example, for an expansion in harmonic oscillator wave functions, the nuclear matrix elements of the multipole operators can be calculated using Ref. \cite{HAXTON2008345}, limiting the interaction of the weak probes to a single nucleon, and neglecting the interaction with many-body clusters of nucleons in the nucleus. 

For a vanishing parameter, the spherical Bessel functions behave as $j_{J}\left(\rho\right)\propto\rho^{J}$ (for $\rho\ll1$). As the momentum transfer $q$ is small for $\beta$-decays, compared to any nuclear property, $\epsilon_{qr}\equiv qR$ is a small parameter. Explicit analysis shows that $\hat{C}_{J},\hat{M}_{J}\propto\epsilon_{qr}^{J}$, while $\hat{L}_{J},\hat{E}_{J}\propto\epsilon_{qr}^{J-1}$.

Further simplification is a result of the parity of the multipole operators. Since the spherical harmonics $Y_{J}$ and $\vec{Y}_{JJ1}^{M}$ have parity of $\left(-\right)^{J}$, the basic parities of the multipole operators (before taking into consideration the nuclear current involved) are $\left(-\right)^{J}$ for the $\hat{C}_{J}$ and $\hat{M}_{J}$ operators, and $\left(-\right)^{J+1}$ for $\hat{L}_{J}$ and $\hat{E}_{J}$. To these, we add the parity of the nuclear currents. While the weak axial charge $J_{0}^{A}$ and the vector current $\vec{J}^{V}$ have a $\frac{\vec{p}}{2m_{N}}$ dependence at their leading order, which leads to an internal negative parity, the weak vector charge $J_{0}^{V}$ and the axial current $\vec{J}^{A}$ have no momentum dependence, leading to an internal positive parity at their leading order. In conclusion, $\hat{C}_{J}^{V}$, $\hat{L}_{J}^{V}$, $\hat{E}_{J}^{V}$ and $\hat{M}_{J}^{A}$ (the superscript $V$ ($A$) denotes multipole operators calculated with the vector (axial) symmetry contribution to the weak nuclear current) have a parity of $\left(-\right)^{J}$, while $\hat{C}_{J}^{A}$, $\hat{L}_{J}^{A}$, $\hat{E}_{J}^{A}$ and $\hat{M}_{J}^{V}$ have a parity of $\left(-\right)^{J+1}$. A detailed non-relativistic expansion of the one-body polar-vector and axial-vector currents in powers of $\epsilon_{\text{NR}}$ can be found in the appendix, along with a detailed derivation of the resulting multipole operators and their accuracy estimation. 

The leading dependence of the nuclear matrix element in the momentum transfer $q$ leads also to a characterization of $\beta$-decays into allowed and forbidden transitions. The allowed transitions are, at their leading order, independent of the momentum transfer $q$, and characterized by no change in the orbital angular momentum $L$, and therefore no change in parity. A Fermi (Gamow-Teller) transition is an allowed transition in which $J$, the change in the total angular momentum of the nucleus, is 0 (1).
Other transitions are suppressed by a $q^{L}$ dependence, with $L\geq1$, and therefore are much slower, and are called, for historical reasons, forbidden transitions. Those are divided into different $L^{\text{th}}$-forbidden transitions, each with a change of orbital angular momentum $L$,
resulting in a parity change of $\left(-\right)^L$.
$\beta$-decays in which the emitted leptons pair have a total spin angular momentum of $S=1$, and an orbital angular momentum $L=J-1$, are called unique $L^{\text{th}}$-forbidden transitions.

The multipole operators depend upon the nuclear model via the nuclear current operators, and connect between the initial (decaying) and final nucleus. This is the source of the $\epsilon_{\text{NM}}$ uncertainty described in sec. \ref{Sec:small_parameters}. We further elaborate on this in sec.~\ref{sec: BSM Signatures}.

\section{A general expression for nuclear shape and recoil corrections\label{sec: general}}

Assuming the Standard Model V-A coupling, a $\beta$-decay transition,
with a $J_{i}^{\pi_{i}}\rightarrow J_{f}^{\pi_{f}}$ angular momentum
and parity change, contains all the integer total angular momentum changes
which uphold the selection rule $\left|J_{i}-J_{f}\right|\leq J\leq J_{i}+J_{f}$,
with the exact parity $\Delta\pi\equiv\pi_{i}\cdot\pi_{f}$:
\begin{align}
\label{eq:sum over theta^J}
\Theta\left(q,\vec{\beta}\cdot\hat{\nu}\right) & =\sum_{J=\left|J_{i}-J_{f}\right|}^{J_{i}+J_{f}}\Theta^{J^{\Delta\pi}}\left(q,\vec{\beta}\cdot\hat{\nu}\right)\text{.}
\end{align}

The next step is to present $\Theta^{J^{\Delta\pi}}\left(q,\vec{\beta}\cdot\hat{\nu}\right)$
for each $J^{\Delta\pi}$. Starting from $J=0$, Fermi
transition ($J^{\Delta\pi}=0^{+}$) expression, including shape and
recoil next-to-leading order (NLO) corrections, can be written as
\begin{align}
\label{eq: 0+}
\Theta^{0^{+}}\left(q,\vec{\beta}\cdot\hat{\nu}\right) & =\left(1+\delta_{1}^{0^{+}}\right)\left[1+a_{\beta\nu}^{0^{+}}\vec{\beta}\cdot\hat{\nu}+b_{\text{F}}^{0^{+}}\frac{m_{e}}{\epsilon}\right]\left|\left\langle \left\Vert \hat{C}_{0}^{V}\right\Vert \right\rangle \right|^{2}\text{,}
\end{align}
with the NLO spectrum-shape correction:
% $1+\tilde{\delta}_{a}$ comes from $\frac{1+\delta_{a}}{1+\delta_{1}}$ and fulfills $\tilde{\delta}_{a}\equiv\delta_{a}-\delta_{1}$.
\begin{eqnarray}
\label{eq: delta, 0+}
\delta_{1}^{0^{+}} & = & -\frac{\epsilon_0}{q} 2\mathfrak{Re}\frac{\left\langle \left\Vert \hat{L}_{0}^{V}\right\Vert \right\rangle }{\left\langle \left\Vert \hat{C}_{0}^{V}\right\Vert \right\rangle }+\mathcal{O}\left(\epsilon_{\text{recoil}}^{2},\epsilon_{qr}^{2}\epsilon_{\text{NR}}^{2}\right)\text{.}
\end{eqnarray}
The multipole operator
$\hat{C}_{0}^{V}\propto1$ is the Fermi leading order, and $\hat{L}_{0}^{V}\propto\epsilon_{\text{recoil}},\epsilon_{qr}\epsilon_{\text{NR}}$
 \footnote{$\epsilon_{\text{recoil}}$ and $\epsilon_{qr}\epsilon_{\text{NR}}$
are about the same order of magnitude for an endpoint of $\approx2\text{MeV}$,
and $\hat{L}_{0}^{V}$ consists of these two forms of terms} is its NLO recoil correction.

One of the parameters playing an important role in beyond the standard
model experimental searches, is the Fierz interference term, which is
the coefficient of the $\frac{m_{e}}{\epsilon}$ term in the differential
distribution of a $\beta$-decay, and can be extracted from
electron energy spectrum measurements. Fierz term vanishes in the
known $V-A$ differential distribution of allowed $\beta$-decay transitions.
Taking into account NLO recoil corrections, Fierz interference term at the Fermi transition, gets the form
% This term is actually multiplied by $1-\delta_{1}$, coming from $\frac{1}{1+\delta_{1}}$, but it is neglected since $\delta_{b}$ is itself already small.
\begin{align}
\label{eq: Fierz, 0+}
b_{\text{F}}^{0^{+}}=\delta_{b}^{0^{+}} & \equiv \frac{m_{e}}{q} 2\mathfrak{Re}\frac{\left\langle \left\Vert \hat{L}_{0}^{V}\right\Vert \right\rangle }{\left\langle \left\Vert \hat{C}_{0}^{V}\right\Vert \right\rangle }+\mathcal{O}\left(\frac{m_{e}}{\epsilon}\epsilon_{\text{recoil}}^{2},\frac{m_{e}}{\epsilon}\epsilon_{qr}^{2}\epsilon_{\text{NR}}^{2}\right)\text{.}
\end{align}

Another parameter of interest for BSM searches, is the
angular correlation between the emitted electron and neutrino,
$a_{\beta\nu}$, which is the coefficient of the $\vec{\beta}\cdot\hat{\nu}$
term in the differential distribution of a $\beta$-decay. The
$V-A$ structure of the weak interaction entails that for a Fermi
transition the $\beta-\nu$ correlation is exactly $a_{\beta\nu}^{0^{+}}=1$.

As for $J^{\Delta\pi}=0^{-}$, which is a non-unique first-forbidden transition,
its $\Theta^{0^{-}}\left(q,\vec{\beta}\cdot\hat{\nu}\right)$ expression takes the exact form
\begin{multline}
\Theta^{0^{-}}\left(q,\vec{\beta}\cdot\hat{\nu}\right)=
\left(1+\vec{\beta}\cdot\hat{\nu}\right)\left[\left|\left\langle \left\Vert \hat{C}_{0}^{A}\right\Vert \right\rangle \right|^{2}+\left|\left\langle \left\Vert \hat{L}_{0}^{A}\right\Vert \right\rangle \right|^{2}-\frac{\epsilon_0}{q}2\mathfrak{Re}\left(\left\langle \left\Vert \hat{L}_{0}^{A}\right\Vert \right\rangle \left\langle \left\Vert \hat{C}_{0}^{A}\right\Vert \right\rangle ^{*}\right)\right]+\\
+\frac{m_{e}^{2}}{q\epsilon}2\mathfrak{Re}\left(\left\langle \left\Vert \hat{L}_{0}^{A}\right\Vert \right\rangle \left\langle \left\Vert \hat{C}_{0}^{A}\right\Vert \right\rangle ^{*}\right)+2\frac{\epsilon\left(\epsilon_0-\epsilon\right)}{q^{2}}\left[\beta^{2}-\left(\vec{\beta}\cdot\hat{\nu}\right)^{2}\right]\left|\left\langle \left\Vert \hat{L}_{0}^{A}\right\Vert \right\rangle \right|^{2}\text{,}\label{eq: 0-}
\end{multline}
with $\hat{C}_{0}^{A}\propto\epsilon_{\text{NR}}$ and $\hat{L}_{0}^{A}\propto\epsilon_{qr}$.
For this $J^{\Delta\pi}=0^{-}$ first-forbidden transition, the angular
correlation coefficient can be easily recognized as
$a_{\beta\nu}^{0^{-}} =1$,
and the Fierz term can be extracted,
\begin{align}
\label{eq: Fierz, 0-}
b_{\text{F}}^{0^{-}}= & \frac{\frac{m_{e}}{q}2\mathfrak{Re}\left(\left\langle \left\Vert \hat{L}_{0}^{A}\right\Vert \right\rangle \left\langle \left\Vert \hat{C}_{0}^{A}\right\Vert \right\rangle ^{*}\right)}{\left|\left\langle \left\Vert \hat{C}_{0}^{A}\right\Vert \right\rangle \right|^{2}+\left|\left\langle \left\Vert \hat{L}_{0}^{A}\right\Vert \right\rangle \right|^{2}-\frac{\epsilon_0}{q}2\mathfrak{Re}\left(\left\langle \left\Vert \hat{L}_{0}^{A}\right\Vert \right\rangle \left\langle \left\Vert \hat{C}_{0}^{A}\right\Vert \right\rangle ^{*}\right)}\text{.}
\end{align}

In order to discuss the $\Theta^{J^{\Delta\pi}}\left(q,\vec{\beta}\cdot\hat{\nu}\right)$
expressions for $J$'s grater than $0$, we will distinguish between
two types of $0<J^{\Delta\pi}$ transitions:
$\Delta\pi=\left(-\right)^{J}$, and $\Delta\pi=\left(-\right)^{J-1}$.
The first type, $J^{\left(-\right)^{J}}$, presents non-unique $J^{\rm th}$ forbidden transitions. The second type, $J^{\left(-\right)^{J-1}}$, presents, for $J=1$, the allowed Gamow-Teller transition, and for $J>1$, unique $(J-1)^{\rm th}$ forbidden transitions (together we will refer to them as unique transitions).

Let us start with the unique transitions (e.g., Gamow\textendash Teller
transition, which is $J^{\Delta\pi}=1^{+}$). 
A general expression,
including shape and recoil NLO corrections, for any unique transition, i.e., a decay with $0<J^{\left(-\right)^{J-1}}$,
in the sum above (eq. \eqref{eq:sum over theta^J}), can be written as:
\begin{align}
    \label{eq: -^(J-1) J>0}
    \Theta^{J^{\left(-\right)^{J-1}}}\left(q,\vec{\beta}\cdot\hat{\nu}\right) & = \frac{2J+1}{J}\left(1+\delta_{1}^{J^{\left(-\right)^{J-1}}}\right)\left\{ 1+a_{\beta\nu}^{J^{\left(-\right)^{J-1}}}\hat{\nu}\cdot\vec{\beta}+b_{\rm F}^{J^{\left(-\right)^{J-1}}}\frac{m_{e}}{\epsilon}+\right.\nonumber\\
    & \left.+\frac{J-1}{2J+1}\frac{\epsilon\left(\epsilon_0-\epsilon\right)}{q^{2}}\left(1+\tilde{\delta}_{\beta^2}^{J^{\left(-\right)^{J-1}}}\right)\left[\vec{\beta}^{2}-\left(\hat{\nu}\cdot\vec{\beta}\right)^{2}\right]\right\}
    \left|\left\langle \left\Vert \hat{L}_{J}^{A}\right\Vert \right\rangle \right|^{2} \text{,}
\end{align}
with the shape and recoil NLO corrections:
\begin{align}
    \label{eq: delta1 J>0}
    \delta_{1}^{J^{\left(-\right)^{J-1}}} & =\frac{2}{2J+1}\mathfrak{Re}\left[-J\frac{\epsilon_0}{q}\frac{\left\langle \left\Vert \hat{C}_{J}^{A}\right\Vert \right\rangle }{\left\langle \left\Vert \hat{L}_{J}^{A}\right\Vert \right\rangle }\pm\sqrt{J\left(J+1\right)}\frac{\epsilon_0-2\epsilon}{q}\frac{\left\langle \left\Vert \hat{M}_{J}^{V}\right\Vert \right\rangle }{\left\langle \left\Vert \hat{L}_{J}^{A}\right\Vert \right\rangle }+\sqrt{J\left(J+1\right)}\frac{\left\langle \left\Vert \hat{E}_{J}^{A\left(\rm res\right)}\right\Vert \right\rangle }{\left\langle \left\Vert \hat{L}_{J}^{A}\right\Vert \right\rangle }\right] + \nonumber \\
    &+\mathcal{O}\left(\epsilon_{qr}^2 \epsilon_{\rm NR}^2 ,\epsilon_{\text{recoil}}^{2}\right),
    \nonumber\\
    \tilde{\delta}_{\beta^2}^{J^{\left(-\right)^{J-1}}} & =-\frac{2}{2J+1}\mathfrak{Re}\left[-J\frac{\epsilon_0}{q}\frac{\left\langle \left\Vert \hat{C}_{J}^{A}\right\Vert \right\rangle }{\left\langle \left\Vert \hat{L}_{J}^{A}\right\Vert \right\rangle }\pm\sqrt{J\left(J+1\right)}\frac{\epsilon_0-2\epsilon}{q}\frac{\left\langle \left\Vert \hat{M}_{J}^{V}\right\Vert \right\rangle }{\left\langle \left\Vert \hat{L}_{J}^{A}\right\Vert \right\rangle }+\frac{3J}{J-1}\sqrt{J\left(J+1\right)}\frac{\left\langle \left\Vert \hat{E}_{J}^{A\left(\rm res\right)}\right\Vert \right\rangle }{\left\langle \left\Vert \hat{L}_{J}^{A}\right\Vert \right\rangle }\right]+ \nonumber \\
    &+\mathcal{O}\left(\epsilon_{qr}^2 \epsilon_{\rm NR}^2 ,\epsilon_{\text{recoil}}^{2}\right) \text{,}
\end{align}
where $\pm$ are for $\beta^{\mp}$ decays. Here the multipole operator $\hat{L}_{J}^{A}\propto\epsilon_{qr}^{J-1}$
is the leading order operator, while $\frac{\hat{C}_{J}^{A}}{\hat{L}_{J}^{A}},\frac{\hat{M}_{J}^{V}}{\hat{L}_{J}^{A}}\propto\epsilon_{\text{recoil}},\epsilon_{qr}\epsilon_{\text{NR}}$
\footnote{Both multipoles, $\hat{C}_{J}^{A}$ and $\hat{M}_{J}^{V}$, consist
of these two forms of terms. $\hat{C}_{J}^{A}$ has an additional
smaller term, proportional to $\epsilon_{m_{\pi}}\equiv\frac{\epsilon_0 q}{m_{\pi}^{2}}$}
are NLO recoil corrections operators.
To eliminate the electric multipole operator $\hat{E}_{J}$, we used
the relation 
\begin{align}
\label{eq: E(res)}
    \hat{E}_{J}=\sqrt{\frac{J+1}{J}}\hat{L}_{J}-i\sqrt{\frac{2J+1}{J}}\int d^{3}rj_{J+1}\left(qr\right)\vec{Y}_{J,J+1,1}\left(\hat{r}\right)\cdot\vec{J}\left(\vec{r}\right) \equiv \sqrt{\frac{J+1}{J}}\hat{L}_{J} + \hat{E}_{J}^{\left(\rm res\right)} \text{,}
\end{align}
that leaves a residual correction $\hat{E}_{J}^{\left(\rm res\right)}$
of
$\hat{E}_{J}$ regarding $\hat{L}_{J}$,
and for the unique transitions is reflected in another NLO correction of $\frac{\hat{E}_{J}^{A\left(\rm res\right)}}{\hat{L}_{J}^{A}} \propto \epsilon_{qr}^{2}$.

The $V-A$ structure of the weak interaction entails, for any specific unique transition,
that the $\beta-\nu$ correlations leading order will be $a_{\beta\nu}^{J^{\left(-\right)^{J-1}}}=-\frac{1}{2J+1}$.
Adding the NLO corrections, the $\beta-\nu$ correlation becomes
\begin{align}
a_{\beta\nu}^{J^{\left(-\right)^{J-1}}} & =-\frac{1}{2J+1}\left(1+\tilde{\delta}_{a}^{J^{\left(-\right)^{J-1}}}\right)\text{,}\label{eq: a_bn, -^(J-1) J>0}
\end{align}
with
\begin{align}
\label{eq: delta_a J>0}
\tilde{\delta}_{a}^{J^{\left(-\right)^{J-1}}} & =\frac{2J}{2J+1}2\mathfrak{Re}\left[\left(J+1\right)\frac{\epsilon_0}{q}\frac{\left\langle \left\Vert \hat{C}_{J}^{A}\right\Vert \right\rangle }{\left\langle \left\Vert \hat{L}_{J}^{A}\right\Vert \right\rangle }\pm\sqrt{J\left(J+1\right)}\frac{\epsilon_0-2\epsilon}{q}\frac{\left\langle \left\Vert \hat{M}_{J}^{V}\right\Vert \right\rangle }{\left\langle \left\Vert \hat{L}_{J}^{A}\right\Vert \right\rangle }+\sqrt{J\left(J+1\right)}\frac{\left\langle \left\Vert \hat{E}_{J}^{A\left(\rm res\right)}\right\Vert \right\rangle }{\left\langle \left\Vert \hat{L}_{J}^{A}\right\Vert \right\rangle }\right]+ \nonumber \\
& +\mathcal{O}\left(\epsilon_{qr}^2 \epsilon_{\rm NR}^2 ,\epsilon_{\text{recoil}}^{2}\right) \text{.}
\end{align}
As for the Fierz term, that vanishes for unique
transitions in the $V-A$ structure of the weak interaction, it now gets the form
\begin{align}
\label{eq: delta_b J>0}
b_{\text{F}}^{J^{\left(-\right)^{J-1}}}=\delta_{b}^{J^{\left(-\right)^{J-1}}} & \equiv\frac{1}{2J+1}\frac{m_{e}}{q}2\mathfrak{Re}\left\{ J\frac{\left\langle \left\Vert \hat{C}_{J}^{A}\right\Vert \right\rangle }{\left\langle \left\Vert \hat{L}_{J}^{A}\right\Vert \right\rangle }\pm\sqrt{J\left(J+1\right)}\frac{\left\langle \left\Vert \hat{M}_{J}^{V}\right\Vert \right\rangle }{\left\langle \left\Vert \hat{L}_{J}^{A}\right\Vert \right\rangle }\right\} +\mathcal{O}\left(\frac{m_{e}}{\epsilon}\epsilon_{qr}^2 \epsilon_{\rm NR}^2 ,\frac{m_{e}}{\epsilon}\epsilon_{\text{recoil}}^{2}\right)\text{.}
\end{align}

For the case of non-unique transitions, i.e., a decay with $0<J^{\left(-\right)^{J}}$,
the $\Theta^{J^{\left(-\right)^{J}}}\left(q,\vec{\beta}\cdot\hat{\nu}\right)$
expression can be written as:
\begin{multline}
\Theta^{J^{\left(-\right)^{J}}}\left(q,\vec{\beta}\cdot\hat{\nu}\right)=
\left\{1+\frac{1}{J}\frac{\epsilon_0^{2}}{q^{2}}\left[1-\left(J+1\right)2\mathfrak{Re}\delta^{J^{\left(-\right)^{J}}}\right]+2\frac{\epsilon_0}{q^{2}}\frac{m_{e}^{2}}{\epsilon}\right\}
\left|\left\langle \left\Vert \hat{C}_{J}^{V}\right\Vert \right\rangle \right|^{2}+\left|\left\langle \left\Vert \hat{M}_{J}^{A}\right\Vert \right\rangle \right|^{2}+\\
\pm\sqrt{\frac{J+1}{J}}\left(\frac{\epsilon_0^{2}}{q^{2}}-2\frac{\epsilon_0\epsilon}{q^{2}}+\frac{\epsilon_0}{q^{2}}\frac{m_{e}^{2}}{\epsilon}\right)2\mathfrak{Re}\left[\left\langle \left\Vert \hat{C}_{J}^{V}\right\Vert \right\rangle \left\langle \left\Vert \hat{M}_{J}^{A}\right\Vert \right\rangle ^{*}\left(1-\delta^{J^{\left(-\right)^{J}}}\right)\right]+\\
+\vec{\beta}\cdot\hat{\nu}\left\{\left[1-\frac{2J+1}{J}\frac{\epsilon_0^{2}}{q^{2}}\left(1-\frac{J+1}{2J+1}2\mathfrak{Re} \delta^{J^{\left(-\right)^{J}}}\right)\right]\left|\left\langle \left\Vert \hat{C}_{J}^{V}\right\Vert \right\rangle \right|^{2}-\left|\left\langle \left\Vert \hat{M}_{J}^{A}\right\Vert \right\rangle \right|^{2}+\right.\\
\left.\mp\sqrt{\frac{J+1}{J}}\left(\frac{\epsilon_0^{2}}{q^{2}}-2\frac{\epsilon_0\epsilon}{q^{2}}\right)2\mathfrak{Re}\left[\left\langle \left\Vert \hat{C}_{J}^{V}\right\Vert \right\rangle \left\langle \left\Vert \hat{M}_{J}^{A}\right\Vert \right\rangle ^{*}\left(1-\delta^{J^{\left(-\right)^{J}}}\right)\right]\right\}+\\
+\frac{\epsilon\left(\epsilon_0-\epsilon\right)}{q^{2}}\left[\beta^{2}-\left(\vec{\beta}\cdot\hat{\nu}\right)^{2}\right]\left[\frac{J-1}{J}\frac{\epsilon_0^{2}}{q^{2}}\left(1+\frac{J+1}{J-1}2\mathfrak{Re}\delta^{J^{\left(-\right)^{J}}}\right)\left|\left\langle \left\Vert \hat{C}_{J}^{V}\right\Vert \right\rangle \right|^{2}-\left|\left\langle \left\Vert \hat{M}_{J}^{A}\right\Vert \right\rangle \right|^{2}\right]\text{,}\label{eq: -^J>0}
\end{multline}
with the NLO correction
\begin{align}
\label{eq: delta, -^J>0}
\delta^{J^{\left(-\right)^{J}}} & =-\frac{q}{\epsilon_0}\sqrt{\frac{J}{J+1}}\frac{\left\langle \left\Vert \hat{E}_{J}^{V\left(\rm res\right)}\right\Vert \right\rangle }{\left\langle \left\Vert \hat{C}_{J}^{V}\right\Vert \right\rangle }+\mathcal{O}\left(\epsilon_{qr}^{2}\epsilon_{\text{NR}}^{2}\right)\text{.}
\end{align}
Using the vector current conservation hypothesis (exact to relevant orders in chiral EFT), which eliminates $\hat{L}_{J}^{V}=\frac{\epsilon_0}{q}\hat{C}_{J}^{V}$
for $J>0$ \cite{Walecka:819209}, the multipole operators involved in the leading order
are only $\hat{C}_{J}^{V},\hat{M}_{J}^{A}\propto\epsilon_{qr}^{J}$, and the NLO operator is  $\frac{\hat{E}_{J}^{V\left(\rm res\right)}}{\hat{C}_{J}^{V}}\propto\epsilon_{qr}\epsilon_{\text{NR}}$
(the vector current $\vec{J}^{V}\left(\vec{r}\right)\propto\epsilon_{\text{NR}}$).
For non-unique forbidden decays, the Fierz term, and the angular correlation coefficient, are not well defined.

\section{Nuclear shape and recoil corrections for allowed transitions \label{sec: allowed}}

For example, let us consider the most general allowed transition, e.g., mixed Fermi
($J^{\Delta\pi}=0^{+}$) and Gamow-Teller ($J^{\Delta\pi}=1^{+}$)
transitions. This requires $\Delta\pi=+$ and $J_{i}=J_{f}>0$,
so, following eq.
\eqref{eq:general-decay-rate} and \eqref{eq:sum over theta^J}, the
decay rate, including NLO corrections, will contain
a sum of $\Theta^{0^{+}}$ (eq. \eqref{eq: 0+}) and $\Theta^{1^{+}}$ (eq. \eqref{eq: -^(J-1) J>0})
%over $J^{\Delta\pi}=0^{+},1^{+}$
(other $\Theta^{J^{+}}$'s, with $J^+$s up to $J_{i}+J_{f}$, also
participate in the sum, but they contribute only higher orders):
\begin{multline}
\frac{d^5\omega}{d\epsilon\frac{d\Omega_{k}}{4\pi}\frac{d\Omega_{\nu}}{4\pi}}=\frac{4}{\pi^{2}}\frac{1}{2J_{i}+1}\left(\epsilon_0-\epsilon\right)^{2}k\epsilon F^{\mp}\left(Z_{f},\epsilon\right)C_{\text{corrections}}\cdot \\
\cdot\left\{ 
\left(1+\delta_{1}^{0^{+}}\right)\left[1+\vec{\beta}\cdot\hat{\nu}
+b_{\text{F}}^{0^{+}}\frac{m_{e}}{\epsilon}\right]\left|\left\langle \left\Vert \hat{C}_{0}^{V}\right\Vert \right\rangle \right|^{2} +3\left(1+\delta_{1}^{1^{+}}\right)\left[1-\frac{1}{3}\vec{\beta}\cdot\hat{\nu}\left(1+\tilde{\delta}_{a}^{1^{+}}\right)
+b_{\text{F}}^{1^{+}}\frac{m_{e}}{\epsilon}
\right]\left|\left\langle \left\Vert \hat{L}_{1}^{A}\right\Vert \right\rangle \right|^{2}\right\} = \\
=\frac{4}{\pi^{2}}\frac{1}{2J_{i}+1}\left(\epsilon_0-\epsilon\right)^{2}k\epsilon F^{\mp}\left(Z_{f},\epsilon\right)C_{\text{corrections}} 
\left(\left|\left\langle \left\Vert \hat{C}_{0}^{V}\right\Vert \right\rangle \right|^{2}+3\left|\left\langle \left\Vert \hat{L}_{1}^{A}\right\Vert \right\rangle \right|^{2}\right)
\cdot\\
\cdot\left(1+\delta_{1}^{\text{F+GT}}\right)\left[1+\vec{\beta}\cdot\hat{\nu}\left(1+\tilde{\delta}_{a}^{\text{F+GT}}\right)\frac{\left|\left\langle \left\Vert \hat{C}_{0}^{V}\right\Vert \right\rangle \right|^{2}-\left|\left\langle \left\Vert \hat{L}_{1}^{A}\right\Vert \right\rangle \right|^{2}}{\left|\left\langle \left\Vert \hat{C}_{0}^{V}\right\Vert \right\rangle \right|^{2}+3\left|\left\langle \left\Vert \hat{L}_{1}^{A}\right\Vert \right\rangle \right|^{2}}
+b_F^{\text{F+GT}}\frac{m_e}{\epsilon}
\right]
\text{,}
\end{multline}
with the NLO corrections:
\begin{align}
\delta_{1}^{\text{F+GT}} & =\frac{\delta_{1}^{0^{+}}\left|\left\langle \left\Vert \hat{C}_{0}^{V}\right\Vert \right\rangle \right|^{2}+3\delta_{1}^{1^{+}}\left|\left\langle \left\Vert \hat{L}_{1}^{A}\right\Vert \right\rangle \right|^{2}}{\left|\left\langle \left\Vert \hat{C}_{0}^{V}\right\Vert \right\rangle \right|^{2}+3\left|\left\langle \left\Vert \hat{L}_{1}^{A}\right\Vert \right\rangle \right|^{2}},\nonumber \\
\tilde{\delta}_{a}^{\text{F+GT}} & =\frac{\delta_1^{0^{+}}\left|\left\langle \left\Vert \hat{C}_{0}^{V}\right\Vert \right\rangle \right|^{2}-\left(\delta_1^{1^{+}}+\tilde{\delta}_{a}^{1^{+}}\right)\left|\left\langle \left\Vert \hat{L}_{1}^{A}\right\Vert \right\rangle \right|^{2}}{\left|\left\langle \left\Vert \hat{C}_{0}^{V}\right\Vert \right\rangle \right|^{2}-\left|\left\langle \left\Vert \hat{L}_{1}^{A}\right\Vert \right\rangle \right|^{2}} \text{,}
\end{align}
where $\delta_{1}^{0^{+}}$ is given in eq. \eqref{eq: delta, 0+}, and $\delta_{1}^{1^{+}}$
and $\tilde{\delta}_{a}^{1^{+}}$ can be found easily from
eq. \eqref{eq: delta1 J>0} and \eqref{eq: delta_a J>0}:
\begin{align}
\delta_{1}^{1+} & =\frac{2}{3} \mathfrak{Re}\left[-\frac{\epsilon_0}{q}\frac{\left\langle \left\Vert \hat{C}_{1}^{A}\right\Vert \right\rangle }{\left\langle \left\Vert \hat{L}_{1}^{A}\right\Vert \right\rangle }\pm\sqrt{2}\frac{\epsilon_0-2\epsilon}{q}\frac{\left\langle \left\Vert \hat{M}_{1}^{V}\right\Vert \right\rangle }{\left\langle \left\Vert \hat{L}_{1}^{A}\right\Vert \right\rangle }+\sqrt{2}\frac{\left\langle \left\Vert \hat{E}_{1}^{A\left(\rm res\right)}\right\Vert \right\rangle }{\left\langle \left\Vert \hat{L}_{1}^{A}\right\Vert \right\rangle }\right] +\mathcal{O}\left(\epsilon_{qr}^2 \epsilon_{\rm NR}^2 ,\epsilon_{\text{recoil}}^{2}\right)\nonumber \\
\tilde{\delta}_{a}^{1+} & =\frac{4}{3}\mathfrak{Re}\left[\frac{2\epsilon_0}{q}\frac{\left\langle \left\Vert \hat{C}_{1}^{A}\right\Vert \right\rangle }{\left\langle \left\Vert \hat{L}_{1}^{A}\right\Vert \right\rangle }\pm\sqrt{2}\frac{\epsilon_0-2\epsilon}{q}\frac{\left\langle \left\Vert \hat{M}_{1}^{V}\right\Vert \right\rangle }{\left\langle \left\Vert \hat{L}_{1}^{A}\right\Vert \right\rangle }+\sqrt{2}\frac{\left\langle \left\Vert \hat{E}_{1}^{A\left(\rm res\right)}\right\Vert \right\rangle }{\left\langle \left\Vert \hat{L}_{1}^{A}\right\Vert \right\rangle }\right] +\mathcal{O}\left(\epsilon_{qr}^2 \epsilon_{\rm NR}^2,\epsilon_{\text{recoil}}^{2}\right)\text{.}
\end{align}
Here the multipole operators $\hat{C}_{0}^{V},\hat{L}_{1}^{A}\propto1$
are the Fermi and Gamow\textendash Teller leading orders, while $\hat{L}_{0}^{V},\hat{C}_{1}^{A},$$\hat{M}_{1}^{V}\propto\epsilon_{\text{recoil}},\epsilon_{qr}\epsilon_{\text{NR}}$ and $\hat{E}_{1}^{A\left(\rm res\right)} \propto \epsilon_{qr}^2$
are their NLO recoil corrections.

As mentioned in section \ref{sec: general}, Fierz term,
which vanishes for $V-A$ allowed decays, does not vanish when taking into
account these corrections. For a pure Fermi transition, its new form
already described in eq. \eqref{eq: Fierz, 0+}, while for a
pure Gamow\textendash Teller, it is (from eq. \eqref{eq: delta_b J>0})
\begin{align}
b_{\text{F}}^{1^{+}}=\delta_b^{1+}=\frac{2m_{e}}{3q}\mathfrak{Re}\left\{ \frac{\left\langle \left\Vert \hat{C}_{1}^{A}\right\Vert \right\rangle }{\left\langle \left\Vert \hat{L}_{1}^{A}\right\Vert \right\rangle }\pm\sqrt{2}\frac{\left\langle \left\Vert \hat{M}_{1}^{V}\right\Vert \right\rangle }{\left\langle \left\Vert \hat{L}_{1}^{A}\right\Vert \right\rangle }\right\} +\mathcal{O}\left(\frac{m_{e}}{\epsilon}\epsilon_{qr}^2 \epsilon_{\rm NR}^2 ,\frac{m_{e}}{\epsilon}\epsilon_{\text{recoil}}^{2}\right).
\end{align}
As for the angular correlation coefficient, it is, as mentioned, exactly 1 for a pure Fermi transition, and gets the corrected form
$a_{\beta\nu}^{1^{+}}=-\frac{1}{3}\left(1+\tilde{\delta}_{a}^{1^{+}}\right)$
for a pure Gamow\textendash Teller (eq. \eqref{eq: a_bn, -^(J-1) J>0}).
Finally, for a mixed Fermi and Gamow\textendash Teller transition one gets:
\begin{align}
a_{\beta\nu}^{\text{F+GT}} & =\frac{\left|\left\langle \left\Vert \hat{C}_{0}^{V}\right\Vert \right\rangle \right|^{2}-\left|\left\langle \left\Vert \hat{L}_{1}^{A}\right\Vert \right\rangle \right|^{2}}{\left|\left\langle \left\Vert \hat{C}_{0}^{V}\right\Vert \right\rangle \right|^{2}+3\left|\left\langle \left\Vert \hat{L}_{1}^{A}\right\Vert \right\rangle \right|^{2}}\left(1+\tilde{\delta}_{a}^{\text{F+GT}}\right),
\nonumber \\
b_{\text{F}}^{\text{F+GT}} & =
\frac{\delta_b^{0+} \left|\left\langle \left\Vert \hat{C}_{0}^{V}\right\Vert \right\rangle \right|^{2}
+3\delta_b^{1+}\left|\left\langle \left\Vert \hat{L}_{1}^{A}\right\Vert \right\rangle \right|^{2}}
{\left|\left\langle \left\Vert \hat{C}_{0}^{V}\right\Vert \right\rangle \right|^{2}+3\left|\left\langle \left\Vert \hat{L}_{1}^{A}\right\Vert \right\rangle \right|^{2}} \text{,}
\end{align}
with $\delta_b^{0+}$ and 
$\delta_b^{1+}$ defined above.

\section{Nuclear shape and recoil corrections for first-forbidden transitions \label{sec: 1fobidden}}

As mentioned at section \ref{sec: formalism}, first-forbidden
transitions are transitions with a $q^{1}$ leading order dependence.
Those will be
transitions involving $J^{\Delta\pi}=0^{-},1^{-},2^{-}$.
Let as now assume the most general first-forbidden transition possible, including
all the first-forbidden $J^{\Delta\pi}$'s. That will be any transition
with $\Delta\pi=-$ and $J_{i}=J_{f}\geq1$ (e.g., the transition $1^{+}\rightarrow1^{-}$).
This way, following eq. \eqref{eq:general-decay-rate} and \eqref{eq:sum over theta^J},
the leading order of its decay rate, with multipoles proportional
to $\epsilon_{\text{NR}}$ and $\epsilon_{qr}$, should include a
sum over $J^{\Delta\pi}=0^{-},1^{-},2^{-}$ (other $J^{-}$'s can
also participate in the sum, up to $J_{i}+J_{f}$, but they will contribute
only higher orders):
\begin{multline}
\frac{d^5\omega}{d\epsilon\frac{d\Omega_{k}}{4\pi}\frac{d\Omega_{\nu}}{4\pi}}=\frac{4}{\pi^{2}}\frac{1}{2J_{i}+1}\left(\epsilon_0-\epsilon\right)^{2}k\epsilon F^{\mp}\left(Z_{f},\epsilon\right)C_{\text{corrections}}\cdot\\
\cdot\left\{ \left(1+\vec{\beta}\cdot\hat{\nu}\right)\left[\left|\left\langle \left\Vert \hat{C}_{0}^{A}\right\Vert \right\rangle \right|^{2}+\left|\left\langle \left\Vert \hat{L}_{0}^{A}\right\Vert \right\rangle \right|^{2}-\frac{\epsilon_0}{q}2\mathfrak{Re}\left(\left\langle \left\Vert \hat{L}_{0}^{A}\right\Vert \right\rangle \left\langle \left\Vert \hat{C}_{0}^{A}\right\Vert \right\rangle ^{*}\right)\right]+\right.\\
+\frac{m_{e}^{2}}{q\epsilon}2\mathfrak{Re}\left(\left\langle \left\Vert \hat{L}_{0}^{A}\right\Vert \right\rangle \left\langle \left\Vert \hat{C}_{0}^{A}\right\Vert \right\rangle ^{*}\right)+2\frac{\epsilon\left(\epsilon_0-\epsilon\right)}{q^{2}}\left(\beta^{2}-\left(\vec{\beta}\cdot\hat{\nu}\right)^{2}\right)\left|\left\langle \left\Vert \hat{L}_{0}^{A}\right\Vert \right\rangle \right|^{2}+\\
+\left(1+\frac{\epsilon_0^{2}}{q^{2}}\left[1-4\mathfrak{Re}\left(\delta^{1^{-}}\right)\right]+2\frac{\epsilon_0}{q^{2}}\frac{m_{e}^{2}}{\epsilon}\right)\left|\left\langle \left\Vert \hat{C}_{1}^{V}\right\Vert \right\rangle \right|^{2}+\left|\left\langle \left\Vert \hat{M}_{1}^{A}\right\Vert \right\rangle \right|^{2}+\\
\pm\sqrt{2}\left(\frac{\epsilon_0^{2}}{q^{2}}-2\frac{\epsilon_0\epsilon}{q^{2}}+\frac{\epsilon_0}{q^{2}}\frac{m_{e}^{2}}{\epsilon}\right)2\mathfrak{Re}\left(\left\langle \left\Vert \hat{C}_{1}^{V}\right\Vert \right\rangle \left\langle \left\Vert \hat{M}_{1}^{A}\right\Vert \right\rangle ^{*}\left(1-\delta^{1^{-}}\right)\right)+\\
+\vec{\beta}\cdot\hat{\nu}\left[\left(1-3\frac{\epsilon_0^{2}}{q^{2}}\left(1-\frac{4}{3}\mathfrak{Re}\delta^{1^{-}}\right)\right)\left|\left\langle \left\Vert \hat{C}_{1}^{V}\right\Vert \right\rangle \right|^{2}-\left|\left\langle \left\Vert \hat{M}_{1}^{A}\right\Vert \right\rangle \right|^{2}+\right.\\
\left.\mp\sqrt{2}\left(\frac{\epsilon_0^{2}}{q^{2}}-2\frac{\epsilon_0\epsilon}{q^{2}}\right)2\mathfrak{Re}\left(\left\langle \left\Vert \hat{C}_{1}^{V}\right\Vert \right\rangle \left\langle \left\Vert \hat{M}_{1}^{A}\right\Vert \right\rangle ^{*}\left(1-\delta^{1^{-}}\right)\right)\right]+\\
-\frac{\epsilon\left(\epsilon_0-\epsilon\right)}{q^{2}}\left[\beta^{2}-\left(\vec{\beta}\cdot\hat{\nu}\right)^{2}\right]\left|\left\langle \left\Vert \hat{M}_{1}^{A}\right\Vert \right\rangle \right|^{2}+\\
\left.
+\frac{5}{2}\left(1+\delta_{1}^{2^-}\right)\left\{ 1-\frac{1}{5}\left(1+\delta_a^{2^-}\right)\hat{\nu}\cdot\vec{\beta}
+\delta_b^{2^-}\frac{m_{e}}{\epsilon}
+\frac{1}{5}\frac{\epsilon\left(\epsilon_0-\epsilon\right)}{q^{2}}\left(1+\tilde{\delta}_{\beta^2}^{2^-}\right)\left[\vec{\beta}^{2}-\left(\hat{\nu}\cdot\vec{\beta}\right)^{2}\right]\right\}
\left|\left\langle \left\Vert \hat{L}_{2}^{A}\right\Vert \right\rangle \right|^{2}
\right\} \text{,}
\end{multline}
%(here we used eq. \eqref{eq: 0-} for $J=0$, \eqref{eq: -^J>0} for $J=1$, and \eqref{eq: -^(J-1) J>0} for $J=2$)
with the shape and recoil
NLO corrections:
%(from eq. \eqref{eq: delta, -^J>0}, \eqref{eq: delta1 J>0} and \eqref{eq: delta_a J>0})
\begin{align}
    \delta^{1^{-}} & =-\frac{1}{\sqrt{2}}\frac{q}{\epsilon_0}\frac{\left\langle \left\Vert \hat{E}_{1}^{V\left(\rm res\right)}\right\Vert \right\rangle }{\left\langle \left\Vert \hat{C}_{1}^{V}\right\Vert \right\rangle }+\mathcal{O}\left(\epsilon_{qr}^{2}\epsilon_{\text{NR}}^{2}\right),\nonumber \\
    \delta_{1}^{2^-} & =\frac{2}{5}\mathfrak{Re}\left[-\frac{2\epsilon_0}{q}\frac{\left\langle \left\Vert \hat{C}_{2}^{A}\right\Vert \right\rangle }{\left\langle \left\Vert \hat{L}_{2}^{A}\right\Vert \right\rangle }\pm\sqrt{6}\frac{\epsilon_0-2\epsilon}{q}\frac{\left\langle \left\Vert \hat{M}_{2}^{V}\right\Vert \right\rangle }{\left\langle \left\Vert \hat{L}_{2}^{A}\right\Vert \right\rangle }
    +\sqrt{6}\frac{\left\langle \left\Vert \hat{E}_{2}^{A\left(\rm res\right)}\right\Vert \right\rangle }{\left\langle \left\Vert \hat{L}_{2}^{A}\right\Vert \right\rangle }\right] 
    +\mathcal{O}\left(\epsilon_{qr}^2 \epsilon_{\rm NR}^2 ,\epsilon_{\text{recoil}}^{2}\right), \nonumber\\
    \tilde{\delta}_{\beta^2}^{2^-} & =-\frac{2}{5}\mathfrak{Re}\left[-\frac{2\epsilon_0}{q}\frac{\left\langle \left\Vert \hat{C}_{2}^{A}\right\Vert \right\rangle }{\left\langle \left\Vert \hat{L}_{2}^{A}\right\Vert \right\rangle }\pm\sqrt{6}\frac{\epsilon_0-2\epsilon}{q}\frac{\left\langle \left\Vert \hat{M}_{2}^{V}\right\Vert \right\rangle }{\left\langle \left\Vert \hat{L}_{2}^{A}\right\Vert \right\rangle }+6\sqrt{6}\frac{\left\langle \left\Vert \hat{E}_{2}^{A\left(\rm res\right)}\right\Vert \right\rangle }{\left\langle \left\Vert \hat{L}_{2}^{A}\right\Vert \right\rangle }\right]+\mathcal{O}\left(\epsilon_{qr}^2 \epsilon_{\rm NR}^2 ,\epsilon_{\text{recoil}}^{2}\right), \nonumber\\
    \tilde{\delta}_{a}^{2^-} & =\frac{8}{5}\mathfrak{Re}\left[\frac{3\epsilon_0}{q}\frac{\left\langle \left\Vert \hat{C}_{2}^{A}\right\Vert \right\rangle }{\left\langle \left\Vert \hat{L}_{2}^{A}\right\Vert \right\rangle }\pm\sqrt{6}\frac{\epsilon_0-2\epsilon}{q}\frac{\left\langle \left\Vert \hat{M}_{2}^{V}\right\Vert \right\rangle }{\left\langle \left\Vert \hat{L}_{2}^{A}\right\Vert \right\rangle }+\sqrt{6}\frac{\left\langle \left\Vert \hat{E}_{2}^{A\left(\rm res\right)}\right\Vert \right\rangle }{\left\langle \left\Vert \hat{L}_{2}^{A}\right\Vert \right\rangle }\right]
    +\mathcal{O}\left(\epsilon_{qr}^2 \epsilon_{\rm NR}^2 ,\epsilon_{\text{recoil}}^{2}\right) \text{.}
\end{align}
Here the multipole operators $\hat{C}_{0}^{A}\propto\epsilon_{\text{NR}}$
and $\hat{L}_{0}^{A},\hat{C}_{1}^{V},\hat{M}_{1}^{A},\hat{L}_{2}^{A}\propto\epsilon_{qr}$
are the first-forbidden leading orders, while $\frac{\hat{E}_{1}^{V\left(\rm res\right)}}{\hat{C}_{1}^{V}}\propto\epsilon_{qr}\epsilon_{\text{NR}}$
, $\frac{\hat{C}_{2}^{A}}{\hat{L}_{2}^{A}},\frac{\hat{M}_{2}^{V}}{\hat{L}_{2}^{A}}\propto\epsilon_{qr}\epsilon_{\text{NR}},\epsilon_{\text{recoil}}$ and 
$\frac{\hat{E}_{2}^{A\left(\rm res\right)}}{\hat{L}_{2}^{A}}\propto\epsilon_{qr}^2$
are their NLO corrections.
As we showed at \cite{GLICKMAGID2017285}, the unique first-forbidden transition, $J^{\Delta\pi}=2^{-}$, is of a great interest for Beyond the Standard Model searches. Its Fierz term, including NLO corrections, will be (from eq. \eqref{eq: delta_b J>0}) $b_{\text{F}}^{2^{-}}=\frac{2m_{e}}{5q}\mathfrak{Re}\left\{ 2 \frac{\left\langle \left\Vert \hat{C}_{2}^{A}\right\Vert \right\rangle }{\left\langle \left\Vert \hat{L}_{2}^{A}\right\Vert \right\rangle }\pm\sqrt{6}\frac{\left\langle \left\Vert \hat{M}_{2}^{V}\right\Vert \right\rangle }{\left\langle \left\Vert \hat{L}_{2}^{A}\right\Vert \right\rangle }\right\} +\mathcal{O}\left(\frac{m_{e}}{\epsilon}\epsilon_{qr}^{2} \epsilon_{\rm NR}^{2},\frac{m_{e}}{\epsilon}\epsilon_{\text{recoil}}^{2}\right)$,
and its angular correlation coefficient will be (eq. \eqref{eq: a_bn, -^(J-1) J>0})
$a_{\beta\nu}^{2^{-}}=-\frac{1}{5}\left(1+\tilde{\delta}_{a}^{2^{-}}\right)$
\footnote{For $J^{\Delta\pi}=0^{-}$ Fierz term and the angular correlation
are in eq. (\ref{eq: Fierz, 0-}) and above it,
while for $J^{\Delta\pi}=1^{-}$ they are not well defined.}.

\section{From nuclear structure corrections to BSM signatures \label{sec: BSM Signatures}}
The matrix elements appearing in the formulae in the previous sections depend upon the nuclear interactions, i.e., the Hamiltonian and the coupling to the weak probe. 
Nuclear interactions are the low-energy reflection of the fundamental QCD forces. Thus, one uses an effective approach to describe the nuclear Hamiltonian and the nuclear currents excited by the weak probe. 
A most common approach is the EFT approach, which systematically builds, order by order, the nuclear Hamiltonian and currents (the scattering
operators), based on the symmetries of the fundamental theory. As EFT creates an expansion in a small parameter, $Q/\Lambda_b$, it is also a source of systematic uncertainty, which we coined $\epsilon_{\text{EFT}}$ in sec. \ref{Sec:small_parameters}, of the order of $\epsilon_{\text{EFT}}^{(n+1)}$, with $n$ the order of the expansion. 

Pragmatically, many calculations use the so called ``impulse'' approximation for the nuclear currents, which describes the interaction of the weak probe with the strongly interacting nucleons, and neglects the interaction of the probe with two or more nucleons simultaneously. In terms of an EFT calculation, this is a leading order approximation,  which entails an ${\mathcal{O}}(\epsilon_{EFT})$ accuracy, usually due to the leading magnetic multipole. This is due to the fact that such currents arise at next-to-leading order for the Magnetic multipole of the polar-vector current. Other significant multipoles receive two-body corrections at higher orders. 
The one-body currents arise consistently at leading order, and their operator structure has been developed already from a phenomenological approach, using Lorentz symmetry considerations, and thus will be the starting point for theoretical predictions. In this section we outline these impulse approximation currents, and analyse their expected accuracy, from the EFT point of view. We emphasize that adding
two-body currents has the potential to significantly increase the accuracy.

The one-body currents include low-energy coefficients, related to the symmetries
of the probe-nucleon interactions. These
are denoted by $C_{V}$ and $C_{A}$ for the polar-vector and axial-vector SM
currents, and $C_{S}$, $C_{P}$ and $C_{T}$ for the scalar, pseudo-scalar
and tensor BSM currents. These symmetry coefficients couple to the nuclear charges $g_{\text{sym}}$
($\text{sym}\in\left\{ S,P,V,A,T\right\} $). 

The vector and axial vector one-body
currents are (respectively)\cite{Cirigliano:2013xha}:
\begin{align}
\label{eq: one-body currents}
\left\langle p\left(p_{p}\right)\left|\bar{u}\gamma_{\mu}d\right|n\left(p_{n}\right)\right\rangle  & =\bar{u}_{p}\left(p_{p}\right)\left[g_{V}\left(q^{2}\right)\gamma_{\mu}-i\frac{\tilde{g}_{T\left(V\right)}\left(q^{2}\right)}{2m_{N}}\sigma_{\mu\nu}q^{\nu}+\frac{\tilde{g}_{S}\left(q^{2}\right)}{2m_{N}}q_{\mu}\right]u_{n}\left(p_{n}\right),\nonumber \\
\left\langle p\left(p_{p}\right)\left|\bar{u}\gamma_{\mu}\gamma_{5}d\right|n\left(p_{n}\right)\right\rangle  & =\bar{u}_{p}\left(p_{p}\right)\left[g_{A}\left(q^{2}\right)\gamma_{\mu}-i\frac{\tilde{g}_{T\left(A\right)}\left(q^{2}\right)}{2m_{N}}\sigma_{\mu\nu}q^{\nu}+\frac{\tilde{g}_{P}\left(q^{2}\right)}{2m_{N}}q_{\mu}\right]\gamma_{5}u_{n}\left(p_{n}\right).
\end{align}
In the Standard Model, $g_{V}=1$, up to second order corrections
in isospin breaking \cite{Ademollo1964,DONOGHUE1990243}, as a result
of the conservation of the vector current, and $g_{A}\approx1.276g_{V}$
\cite{Mendenhall2013,Mund2013}. $\tilde{g}_{S}$ and $\tilde{g}_{T\left(A\right)}$,
known as second class currents, do not exist in the Standard Model,
$\tilde{g}_{S}$ due to current conservation, and $\tilde{g}_{T\left(A\right)}$
because of G-parity considerations. The scalar, pseudo-scalar and tensor
currents are (respectively)\cite{Cirigliano:2013xha}:
\begin{eqnarray}
\left\langle p\left(p_{p}\right)\left|\bar{u}d\right|n\left(p_{n}\right)\right\rangle  & = & g_{S}\left(0\right)\bar{u}_{p}\left(p_{p}\right)u_{n}\left(p_{n}\right)+\mathcal{O}\left(\epsilon_{\text{recoil}}^{2}\right),\nonumber \\
\left\langle p\left(p_{p}\right)\left|\bar{u}\gamma_{5}d\right|n\left(p_{n}\right)\right\rangle  & = & g_{P}\left(0\right)\bar{u}_{p}\left(p_{p}\right)\gamma_{5}u_{n}\left(p_{n}\right)+\mathcal{O}\left(\epsilon_{\text{recoil}}^{2}\right),\nonumber \\
\left\langle p\left(p_{p}\right)\left|\bar{u}\sigma_{\mu\nu}d\right|n\left(p_{n}\right)\right\rangle  & = & g_{T}\left(0\right)\bar{u}_{p}\left(p_{p}\right)\sigma_{\mu\nu}u_{n}\left(p_{n}\right)+\mathcal{O}\left(\epsilon_{\text{recoil}}^{2}\right)\text{,}
\end{eqnarray}
where the scalar nuclear charge, $g_{S}=g_{V}\frac{\left(M_{n}-M_{p}\right)^{QCD}}{m_{d}-m_{u}}\approx0.8-1.2$,
the pseudo-scalar one is $g_{P}=g_{V}\frac{M_{n}+M_{p}}{m_{d}+m_{u}}=349(9)$
(the pseudo-scalar contraction, $\bar{u}_{p}\gamma_{5}u_{n}$, is $\mathcal{O}\left(\epsilon_{\text{recoil}}\right)$,
as opposed to the others, which are $\mathcal{O}\left(1\right)$\cite{Cirigliano:2013xha}),
and according to lattice QCD, the tensor nuclear charge $g_{T}$ has
the same order of magnitude as $g_{A}$\cite{Bhattacharya2016}. Consequently, 
all these nuclear charges have the same orders of magnitude.
The suppression of the symmetry coefficients results from the effective
theory\textquoteright s coefficients, $\epsilon_{\text{sym}}\propto\left(\frac{m_{W}}{\Lambda}\right)^{n}$.
These originate in the effective weak interaction Lagrangian, where
$\Lambda$ represents new physics scale, and $n=0$ for $\text{sym}=V,A$,
and $n\geq2$ for $\text{sym}\neq V,A$. Since it is an effective
theory, it will be surprising if it is accurate and contains only
vector and axial currents. New experiments will have a per-mill level
precision, sensitive to new physics at the TeV scale. For the simplest
BSM operator ($n=2$), a TeV scale means $\epsilon_{\text{sym}}\sim10^{-3}$,
so the needed accuracy of the calculation is about $10^{-3}-10^{-4}$.

As we have shown, the presented formalism reaches an accuracy of at
least $\mathcal{O}\left(\epsilon_{qr}^{2}\epsilon_{\rm NR}^{2}\right)$ or $\mathcal{O}\left(\epsilon_{\text{recoil}}^{2}\right)$.
For an endpoint of $\approx2\text{MeV}$, $\epsilon_{qr}\approx0.01A^{\frac{1}{3}}$, $\epsilon_{NR}\approx0.2$
and $\epsilon_{\text{recoil}}\approx0.002$, this entails a an order of 10\% accuracy for the solution of the nuclear many bady problem ($\epsilon_{\text{solver}}$).

\section{Summary \label{sec: summary}}

In this paper, we present a formalism to calculate $\beta$-decay rates and observables, 
including high order recoil and shape corrections, required for on going and planned experiments
in search for BSM physics, which aim at a per-mill level of accuracy.
The formalism shows that nuclear structure corrections induce a finite Fierz interference term, a fact which affects the analysis of measurements of the angular correlation between the beta particle and the (anti-)neutrino. In addition, We express the nuclear structure related corrections to the angular correlation
coefficient. Both Fierz term and the angular correlation are of interest for ongoing experiments, and thus these corrections are essential
in order to distinguish BSM signatures from high orders effects within
the Standard Model.

We identify different small parameters involved in the corrections, that allow us to robustly assess the accuracy of the theoretical calculations. 
These expansion parameters originate in the low-energy character of $\beta$-decays, the non-relativistic character of nuclear wave function, and the coulomb interaction between the electron and the nucleus, defined as follows:
\begin{eqnarray}
\epsilon_{qr}&\equiv& qR, \nonumber \\
\epsilon_{\text{NR}} &\equiv&\frac{P_{\text{Fermi}}}{m_{N}}, \nonumber\\
\epsilon_{\text{recoil}} &\equiv& \frac{q}{m_{N}}, \nonumber\\
\epsilon_c&\equiv& \alpha Z_{f}.
\end{eqnarray}
In addition we define parameters that characterize the precision of the solution of the nuclear problem, $\epsilon_{\rm NM}$, which contains the EFT expansion order $\epsilon_{\text{EFT}}=Q/\Lambda_b$, its specific application, i.e., the details related to regularization of the effective theory, the non-relativistic expansion, etc. ($\epsilon_{\text{model}}$), and the numerical accuracy $\epsilon_{\text{conv}}$.
These enable an analysis of the accuracy of the calculations, and demonstrate
that a solution of the nuclear many body problem accurate only to about 10\%, can be used to significantly constrain these corrections, for the needs of future experiments.

The application of the presented formalism to specific nuclei is already underway.
The first one is $^6$He \cite{glickmagid2021nuclear}, following its role in several ongoing, or soon to be initiated, precision $\beta$-decay experiments.
The second is $^{23}$Ne \cite{Mishnayot-23Ne}, 
in which a joint theoretical-experimental effort enabled us to reanalyze experimental data using new branching ratio measurements, and establish bounds on the presence of exotic tensor interactions.

\begin{acknowledgments}
   We would like to thank the discussions in the ECT* workshop ``Precise beta decay calculations for searches for new physics'' in Trento, and especially Leendert Hayen, for helpful comments.
   We wish to acknowledge the support of the Israel Science Foundation grant no.\ 1446/16. AGM’s research was partially supported by a scholarship sponsored by the Ministry of Science \& Technology, Israel.
\end{acknowledgments}

\appendix

\section*{Appendix: nuclear currents and multipole operators at leading orders\label{sec: appendix}}

To be able to discuss the nuclear dependent part, we need to look at the nuclear current. In the traditional nuclear physics picture, the electroweak current is constructed from the properties of free nucleons. In the leading order, we can refer to the nuclear currents as one body currents. Neglecting the second-class currents, we rewrite eq.~\eqref{eq: one-body currents} in a more detailed form:
\begin{align}
\left\langle \vec{p}',\sigma',\rho'\left|J_{\mu}^{V}\left(0\right)\right|\vec{p},\sigma,\rho\right\rangle  & =\bar{u}\left(\vec{p'},\sigma'\right)\eta_{\rho'}^{+}\left[g_{V}\left(q^{2}\right)\gamma_{\mu}-i\frac{\tilde{g}_{T\left(V\right)}\left(q^{2}\right)}{2m_{N}}\sigma_{\mu\nu}q^{\nu}\right]\tau^{\pm}\eta_{\rho}u\left(\vec{p},\sigma\right),\nonumber \\
\left\langle \vec{p}',\sigma',\rho'\left|J_{\mu}^{A}\left(0\right)\right|\vec{p},\sigma,\rho\right\rangle  & =\bar{u}\left(\vec{p'},\sigma'\right)\eta_{\rho'}^{+}\left[g_{A}\left(q^{2}\right)\gamma_{\mu}+\frac{\tilde{g}_{P}\left(q^{2}\right)}{2m_{N}}q_{\mu}\right]\gamma_{5}\tau^{\pm}\eta_{\rho}u\left(\vec{p},\sigma\right).
\end{align}
Here $u\left(\vec{p},\sigma\right)=\sqrt{\frac{\epsilon_{p}+m_{N}}{2\epsilon_{p}}}\left(\begin{array}{c}
1\\
\frac{\vec{\sigma}\cdot\vec{p}}{\epsilon_{p}+m_{N}}
\end{array}\right)\chi_{\sigma}$ is Dirac spinor for a free nucleon of mass $m_{N}$, $\epsilon_{p}=\sqrt{p^{2}+m_{N}^{2}}$
is the energy of the particle, $\chi_{\sigma}$ is two-component Pauli
spinor for a spin up and down along the $q$ axis, $\eta_{\rho}$ are
two-component Pauli isospinors, and $\tau^{\pm}=\mp\frac{1}{2}\left(\tau_{x}\pm i\tau_{y}\right)$
are the isospin raising and lowering operators, that turns a proton
into neutron and vice versa.

After substituting the mentioned explicit form of Dirac spinors (using
the convention $\bar{u}=u^{+}\gamma_{0}$, so that $u^{+}u=1$), we
make a non-relativistic expansion, expanding the matrix element consistently
in powers of $\epsilon_{\text{NR}}\equiv\frac{P_{\text{fermi}}}{m_{N}}$,
as momenta are assumed here up to few hundred $\mbox{MeV}$'s, and
find the required matrix elements (here $P_{\mu}=p_{\mu}+p_{\mu}^{'}$,
$q_{\mu}=p_{\mu}-p_{\mu}^{'}$ and $\epsilon_0\equiv q_{0}$):
\begin{align}
\label{eq: Hadron currents}
\left\langle \vec{p}',\sigma',\rho'\left|J_{0}^{V}\left(0\right)\right|\vec{p},\sigma,\rho\right\rangle &=g_{V}\chi_{\sigma'}^{+}\eta_{\rho'}^{+}\tau^{\pm}\eta_{\rho}\chi_{\sigma}+\mathcal{O}\mathrm{\left(\epsilon_{\text{NR}}^{2}\right)},\nonumber \\
\left\langle \vec{p}',\sigma',\rho'\left|\vec{J}^{V}\left(0\right)\right|\vec{p},\sigma,\rho\right\rangle &=\chi_{\sigma'}^{+}\eta_{\rho'}^{+}\frac{1}{2m_{N}}\left[g_{V}\vec{P}+\left(g_{V}+\tilde{g}_{T\left(V\right)}\right)i\vec{q}\times\vec{\sigma}\right]\tau^{\pm}\eta_{\rho}\chi_{\sigma}+\mathcal{O}\mathrm{\left(\epsilon_{\text{NR}}^{2}\right)}, \nonumber \\
\left\langle \vec{p}',\sigma',\rho'\left|J_{0}^{A}\left(0\right)\right|\vec{p},\sigma,\rho\right\rangle &=\chi_{\sigma'}\eta_{\rho'}^{+}\frac{1}{2m_{N}}\left[g_{A}\vec{P}\cdot\vec{\sigma}+\frac{\tilde{g}_{P}}{2m_{N}}\epsilon_0\vec{q}\cdot\vec{\sigma}\right]\tau^{\pm}\eta_{\rho}\chi_{\sigma}+\mathcal{O}\mathrm{\left(\epsilon_{\text{NR}}^{2}\right)}\nonumber, \\
\left\langle \vec{p}',\sigma',\rho'\left|\vec{J}^{A}\left(0\right)\right|\vec{p},\sigma,\rho\right\rangle &=g_{A}\chi_{\sigma'}\eta_{\rho'}^{+}\vec{\sigma}\tau^{\pm}\eta_{\rho}\chi_{\sigma}+\mathcal{O}\mathrm{\left(\epsilon_{\text{NR}}^{2}\right)}\text{.}
\end{align}
Then, we use the definition of the (second quantization) $J\left(\vec{x}\right)$
current matrix element as a sum over first quantization currents $\hat{\mathcal{J}}^{\left(1\right)}$: $\left\langle \vec{p}',\sigma',\rho'\left|J\left(\vec{r}\right)\right|\vec{p},\sigma,\rho\right\rangle =\int d^{3}y\phi_{\vec{p}'\sigma'\rho'}^{+}\left(\vec{y}\right)\left[\hat{\mathcal{J}}^{\left(1\right)}\left(\vec{y}\right)\delta^{\left(3\right)}\left(\vec{r}-\vec{y}\right)\right]\phi_{\vec{p}\sigma\rho}\left(\vec{y}\right)$.
Evaluated at $\vec{r}=0$, we find out that $\left\langle \vec{p}',\sigma',\rho'\left|J\left(0\right)\right|\vec{p},\sigma,\rho\right\rangle =\phi_{\vec{p}'\sigma'\rho'}^{+}\left(0\right)\hat{\mathcal{J}}^{\left(1\right)}\left(0\right)\phi_{\vec{p}\sigma\rho}\left(0\right)$,
what permits the identification of the nuclear density operators in
first quantization (from eq. \eqref{eq: Hadron currents}). Finally,
using the current density operator in the first quantization,
$J\left(\vec{r}\right) =  \sum_{j=1}^{A}\hat{\mathcal{J}}^{\left(1\right)}\left(j\right)\delta^{\left(3\right)}\left(\vec{r}-\vec{r}_{j}\right)$,
and under the first quantization's assumption that there is no dependency
on the location, so $\hat{\mathcal{J}}^{\left(1\right)}\left(j\right)=\hat{\mathcal{J}}^{\left(1\right)}\left(0\right)\left(j\right)$,
one gets the following currents:
\begin{align}
\label{eq: V-A nuclear currents}
J_{0}^{V}\left(\vec{r}\right) & =g_{V}\sum_{j=1}^{A}\tau_{j}^{\pm}\delta^{\left(3\right)}\left(\vec{r}-\vec{r}_{j}\right)+\mathcal{O}\mathrm{\left(\epsilon_{\text{NR}}^{2}\right)},\nonumber \\
\vec{J}^{V}\left(\vec{r}\right) & =\frac{1}{2m_{N}}\sum_{j=1}^{A}\left[g_{V}\left\{ \vec{p}_{j},\delta^{\left(3\right)}\left(\vec{r}-\vec{r}_{j}\right)\right\} +\left(g_{V}+\tilde{g}_{T\left(V\right)}\right)\vec{\nabla}\times\vec{\sigma}_{j}\delta^{\left(3\right)}\left(\vec{r}-\vec{r}_{j}\right)\right]\tau_{j}^{\pm}+\mathcal{O}\mathrm{\left(\epsilon_{\text{NR}}^{2}\right)},\nonumber \\
J_{0}^{A}\left(\vec{r}\right) & =\frac{1}{2m_{N}}\sum_{j=1}^{A}\left[g_{A}\left\{ \vec{p}_{j},\delta^{\left(3\right)}\left(\vec{r}-\vec{r}_{j}\right)\right\} -i\frac{\tilde{g}_{P}}{2m_{N}}\left(\epsilon_0\pm\Delta E_c\right)\vec{\nabla}\delta^{\left(3\right)}\left(\vec{r}-\vec{r}_{j}\right)\right]\cdot\vec{\sigma}_{j}\tau_{j}^{\pm}+\mathcal{O}\mathrm{\left(\epsilon_{\text{NR}}^{2}\right)},\nonumber \\
\vec{J}^{A}\left(\vec{r}\right) & =g_{A}\sum_{j=1}^{A}\vec{\sigma}_{j}\tau_{j}^{\pm}\delta^{\left(3\right)}\left(\vec{r}-\vec{r}_{j}\right)+\mathcal{O}\mathrm{\left(\epsilon_{\text{NR}}^{2}\right)}.
\end{align}
Here we made the operator replacements $\vec{P}\rightarrow\left\{ \vec{p}_{j},\delta^{\left(3\right)}\left(\vec{r}-\vec{r}_{j}\right)\right\} $
, and $\vec{q}\rightarrow-i\vec{\nabla}$, the last one based on a
partial integration of Fourier transform of the transition matrix
element of the current, $\int e^{-i\vec{q}\cdot\vec{r}}\left\langle f\left|J_{\mu}\left(\vec{r}\right)\right|i\right\rangle $,
with localized densities.
We also corrected here the maximal electron energy $\epsilon_0$ with the Coulomb replacement energy $\Delta E_c$, as mentioned in sec. \ref{Sec:small_parameters}.

Positioning eq. \eqref{eq: V-A nuclear currents} into the multipole
operators definition (eq. \eqref{eq: multipole operators}), leads to the explicit expressions for the vector
and axial currents multipole operators:
\begin{align}
\hat{C}_{J}^{V}\left(q\right) & =g_{V}\sum_{j=1}^{A}M_{J}\left(q\vec{r}_{j}\right)\tau_{j}^{\pm}+\mathcal{O}\left(\epsilon_{qr}^{J}\epsilon_{\text{NR}}^{2}\right),\nonumber \\
\hat{L}_{J}^{V}\left(q\right) & =-\frac{q}{2m_{N}}g_{V}\sum_{j=1}^{A}\left\{ M_{J}\left(q\vec{r}_{j}\right)-2\left[\frac{1}{q}\vec{\nabla}M_{J}\left(q\vec{r}_{j}\right)\right]\cdot\frac{1}{q}\vec{\nabla}\right\} \tau_{j}^{\pm}+\mathcal{O}\left(\epsilon_{qr}^{J-1}\epsilon_{\text{NR}}^{2}\right),\nonumber \\
\hat{E}_{J}^{V}\left(q\right) & =\frac{q}{m_{N}}\sum_{j=1}^{A}\left\{ -ig_{V}\left[\frac{1}{q}\vec{\nabla}\times\vec{M}_{JJ1}\left(q\vec{r}_{j}\right)\right]\cdot\frac{1}{q}\vec{\nabla}+\frac{g_{V}+\tilde{g}_{T\left(V\right)}}{2}\vec{M}_{JJ1}\left(q\vec{r}_{j}\right)\cdot\vec{\sigma}_j\right\} \tau_{j}^{\pm}+\mathcal{O}\left(\epsilon_{qr}^{J-1}\epsilon_{\text{NR}}^{2}\right),\nonumber \\
\hat{M}_{J}^{V}\left(q\right) & =-\frac{iq}{m_{N}}\sum_{j=1}^{A}\left\{ g_{V}\vec{M}_{JJ1}\left(q\vec{r}_{j}\right)\cdot\frac{1}{q}\vec{\nabla}+i\frac{g_{V}+\tilde{g}_{T\left(V\right)}}{2}\left[\frac{1}{q}\vec{\nabla}\times\vec{M}_{JJ1}\left(q\vec{r}_{j}\right)\right]\cdot\vec{\sigma}_j\right\} \tau_{j}^{\pm}+\mathcal{O}\left(\epsilon_{qr}^{J}\epsilon_{\text{NR}}^{2}\right)
,\nonumber \\
\hat{C}_{J}^{A}\left(q\right) & =-\frac{iq}{m_{N}}\sum_{j=1}^{A}\left\{ g_{A}M_{J}\left(q\vec{r}_{j}\right)\vec{\sigma}_j\cdot\frac{1}{q}\vec{\nabla}+\frac{1}{2}\left[g_{A}-\frac{\tilde{g}_{P}}{2m_{N}}\left(\epsilon_0\pm\Delta E_c\right)\right]\left[\frac{1}{q}\vec{\nabla}M_{J}\left(q\vec{r}_j\right)\right]\cdot\vec{\sigma}_j\right\} \tau_{j}^{\pm}+\mathcal{O}\left(\epsilon_{qr}^{J}\epsilon_{\text{NR}}^{2}\right),\nonumber \\
\hat{L}_{J}^{A}\left(q\right) & =ig_{A}\sum_{j=1}^{A}\left[\frac{1}{q}\vec{\nabla}M_{J}\left(q\vec{r}_{j}\right)\right]\cdot\vec{\sigma}_j\tau_{j}^{\pm}+\mathcal{O}\left(\epsilon_{qr}^{J-1}\epsilon_{\text{NR}}^{2}\right),\nonumber \\
\hat{E}_{J}^{A}\left(q\right) & =g_{A}\sum_{j=1}^{A}\left[\frac{1}{q}\vec{\nabla}\times\vec{M}_{JJ1}\left(q\vec{r}_{j}\right)\right]\cdot\vec{\sigma}_j\tau_{j}^{\pm}+\mathcal{O}\left(\epsilon_{qr}^{J-1}\epsilon_{\text{NR}}^{2}\right),\nonumber \\
\hat{M}_{J}^{A}\left(q\right) & =g_{A}\sum_{j=1}^{A}\vec{M}_{JJ1}\left(q\vec{r}_{j}\right)\cdot\vec{\sigma}_j\tau_{j}^{\pm}+\mathcal{O}\left(\epsilon_{qr}^{J}\epsilon_{\text{NR}}^{2}\right)\text{,}
\end{align}
where
\begin{eqnarray}
M_{J}\left(q\vec{r}\right) & \equiv & j_{J}\left(qr\right)Y_{J}\left(\hat{r}\right),\nonumber \\
\vec{M}_{JL1}\left(q\vec{r}\right) & \equiv & j_{L}\left(qr\right)\vec{Y}_{JL1}\left(\hat{r}\right)\mbox{,}
\end{eqnarray}
which hold to the identities \cite{1975mpwi.conf..114W}:
\begin{align}
\frac{1}{q}\vec{\nabla}M_{J}\left(q\vec{r}\right) & =\sqrt{\frac{J+1}{2J+1}}\vec{M}_{J,J+1,1}\left(q\vec{r}\right)+\sqrt{\frac{J}{2J+1}}\vec{M}_{J,J-1,1}\left(q\vec{r}\right),\nonumber \\
\frac{1}{q}\vec{\nabla}\times\vec{M}_{JJ1}\left(q\vec{r}\right) & =-i\sqrt{\frac{J}{2J+1}}\vec{M}_{J,J+1,1}\left(q\vec{r}\right)+i\sqrt{\frac{J+1}{2J+1}}\vec{M}_{J,J-1,1}\left(q\vec{r}\right)\text{.}
\end{align}

\bibliographystyle{unsrt}
\addcontentsline{toc}{section}{\refname}\bibliography{bib}

\end{document}